# Limit-cycle-based control of the myogenic wingbeat rhythm in the fruit fly *Drosophila*

Jan Bartussek[1], A. Kadir Mutlu[1], Martin Zapotocky[2,3], Steven N. Fry[4]

[1] Institute of Neuroinformatics, University of Zurich and ETH Zurich, Winterthurerstrasse 190, 8057 Zurich, Switzerland

[2] Institute of Physiology, Academy of Sciences of the Czech Republic, Videnska 1083, 14220 Prague 4, Czech Republic

[3] Institute of Biophysics and Informatics, First Faculty of Medicine, Charles University, Salmovska 1, 12000 Prague 2, Czech Republic

[4] SciTrackS GmbH, Lohzelgstrasse 7, 8118 Pfaffhausen, Switzerland

*Author for correspondence: Jan Bartussek, e-mail: jan@ini.phys.ethz.ch*

**Abstract**

In many animals, rhythmic motor activity is governed by neural limit cycle oscillations under the control of sensory feedback. In the fruit fly Drosophila melanogaster, the wingbeat rhythm is generated myogenically by stretch-activated muscles and hence independently from direct neural input. In this study, we explored if generation and cycle-by-cycle control of Drosophila's wingbeat are functionally separated, or if the steering muscles instead couple into the myogenic rhythm as a weak forcing of a limit cycle oscillator. We behaviourally tested tethered flying flies for characteristic properties of limit cycle oscillators. To this end, we mechanically stimulated the fly's 'gyroscopic' organs, the halteres, and determined the phase relationship between the wing motion and stimulus. The flies synchronized with the stimulus for specific ranges of stimulus amplitude and frequency, revealing the characteristic Arnol´d tongues of a forced limit cycle oscillator. Rapid periodic modulation of the wingbeat frequency prior to locking demonstrates the involvement of the fast steering muscles in the observed control of the wingbeat frequency. We propose that the mechanical forcing of a myogenic limit cycle oscillator permits flies to avoid the comparatively slow control based on a neural central pattern generator.

1. Introduction

Locomotion plays a profound role in essential animal behaviours, such as foraging for food, finding mates and evading predation. In most cases, animals rely on the rhythmic actuation of appendages, such as legs, fins and wings for locomotion. While providing reaction forces for forward propulsion or to remain airborne, these same structures typically also serve as control surfaces to achieve stability of motion, as well as perform impressive manoeuvres.

In many animals, the rhythmic activity underlying locomotion originates in neural central pattern generators (CPGs), whose cyclic activity is modulated by reafferent sensory feedback. CPGs represent a ubiquitous neural mechanism by which animals generate and control rhythmic activity for various bodily functions, such as respiration, chewing and limb actuation [1]. CPGs have been successfully modelled using oscillators with a stable limit cycle (limit cycle oscillators, LCOs) [2-4]. The intrinsic properties of such nonlinear oscillators are illustrated in Fig. 1 by example of the Van der Pol oscillator [5]. The oscillation of position in time (black trace in Fig. 1A) represents a limit cycle orbit in the position-velocity phase space (black curve in Fig. 1B). Limit cycle orbits are asymptotically stable, i.e. the system converges to the limit cycle following a transient perturbation. In the example shown, the delivery of pulse perturbations during three consecutive cycles (Fig. 1A) temporarily affects the periodic activity (green curve), which subsequently returns to its stable limit cycle (Fig. 1B). Depending on the cycle phase, in which such a forcing is applied, the frequency of the LCO is transiently affected [6]. In the present case, the delivery of the pulses leads to a frequency increase (compare green and dashed lines in Fig. 1 A).

In such a way, precisely timed input can control the rhythmic pattern of LCOs efficiently and rapidly – a phenomenon that has also been experimentally demonstrated in CPGs [7, 8]. Computational modelling has shown that phase-dependent reafferent feedback from the periphery to the CPG can ensure robustness of the rhythm in presence of environmental perturbations [9-11]. Such limit-cycle-based control therefore provides a conceptual framework in which to understand how animals generate and control complex motor patterns robustly and efficiently [12].

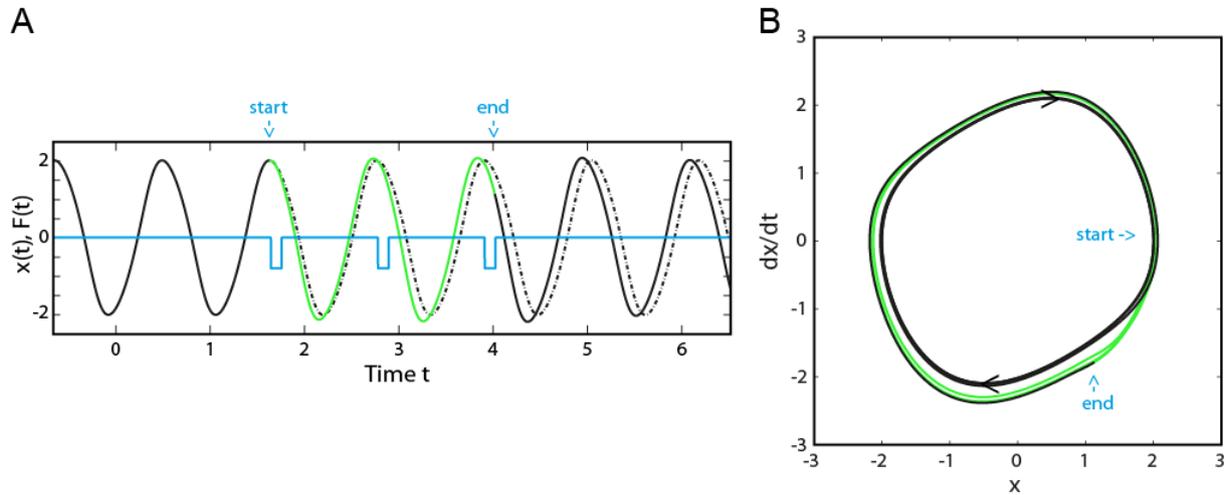

**Figure 1. A) Time course** $x(t)$ **of a van der Pol oscillator perturbed by external force** $F(t)$. The equation of motion is $\ddot{x} - \mu(1-x^2)\dot{x} + x = F(t)$ with $\mu = 0.3$. Blue trace: perturbation $F(t)$. Black trace: $x(t)$ before and after perturbation. Green trace: $x(t)$ during perturbation. Dashed trace: continuation of unperturbed oscillation (perturbation not switched on, $F(t) = 0$). **B) Limit cycle of van der Pol oscillator.** The black and green traces from A) are replotted in the phase space (velocity $dx/dt$ vs. coordinate $x$); increasing time $t$ corresponds to clockwise rotation.

CPGs were originally discovered in locusts [13], in which CPGs provide the neural rhythm for the activity of the thoracic flight muscles giving rise to the wing motion. Synaptic input from the sensory system modulates these patterns to achieve flight control [14-16]. An entirely different actuation mechanism is found in flies, whose highly specialized flight apparatus likewise serves as a model to explore fundamental control mechanisms. The sophisticated flight control abilities of flies are exemplified by the tiny fruit fly *Drosophila melanogaster*, which performs sharp turning manoeuvres within a fraction of a second (around 50 milliseconds, or 10 wingbeats) [17]. Unlike in locusts, the fast and power demanding wingbeat of fruit flies arises myogenically, without direct neural control [18]: Antagonistic sets of stretch activated muscles bring the thorax and with it the wings into resonant oscillation at around 200 Hz. At the time scale of single wing strokes, the wing motion is finely modulated by minuscule steering muscles, which insert directly at the sclerites of the wing articulation and are under immediate neural control [19-22]. The activity of these direct steering muscles depends on a reflexive feedback loop involving specialized mechanosensory organs, the halteres, which sense the rotational velocity of the fly's body [23-29]. According to the current understanding, the powerful, but "dumb" [30] stretch activated muscles supply a periodic force to drive the roughly sinusoidal wing motion [31, 32], while the weak, but fast direct steering muscles modulate stroke position and amplitude to stabilize flight and manoeuvre.

In this way, the highly differentiated muscle types of flies are assumed to bring about the generation and control of the wingbeat rhythm as functionally separate processes [30, 33] - unlike in locusts, in which actuation and control are closely integrated within the framework of a LCO forced by afferent neural input. In the fly, it remains unclear if and how the myogenic rhythm, which cannot be controlled on a cycle-by-cycle basis through direct neuronal input, may be efficiently regulated by sensory feedback. While cycle-by-cycle control of wing stroke amplitude, stroke deviation and mean stroke position by the direct action of the steering muscles has been extensively studied [33-36], the fast control of the thorax/wing oscillator frequency has not been investigated in the previous literature. To achieve such frequency control, the myogenic rhythm would have to be coupled to the only system that is able to act on such fast time scales, i.e., the steering muscles. The null hypothesis of a functional separation between power and steering muscles predicts that flies are unable to control their wingbeat frequency at the time scale of single wing strokes.

While flies apparently do not rely on CPGs to provide their wingbeat rhythm, it is intriguing to consider the possibility that the indirect muscle actuation mechanism itself represents a LCO, which is forced in a phase-dependent manner by the

*mechanical* activity of the steering muscles. This alternative hypothesis is consistent with the "limit-cycle control" scheme for insect flight suggested in [37, 38]. If confirmed, the flight control strategy of flies, and the functional role of their steering muscles in particular, would appear in a new light altogether.

We explored this possibility by testing fruit flies for a functional coupling of the steering muscles onto the wing stroke pattern generated by the stretch activated muscles. We evoked steering muscle activity from a periodic mechanic stimulation of the halteres, which are known to provide direct input to the steering muscles [21]. We used a laser vibrometer to measure stroke-by-stroke variations in the wingbeat frequency of tethered flying flies and attribute these changes unambiguously to the activity of the direct steering muscles, which are the only muscles in the flight apparatus of flies that are capable of responding at the time scale of a single wing stroke.

In the case of sustained periodic forcing, a LCO may adjust its frequency and eventually synchronize with the forcing, a phenomenon that has been observed in various biological limit cycle oscillators (e.g. [39, 40], review [41]). Depending on the value of the forcing frequency and amplitude, one can expect to obtain *1:1* synchronization (entrainment), or a regime in which the forcing frequency and the LCO frequency are other rational multiples of each other, such as *2:1*, *3:2*, etc.. Consistent with this generic property of periodically forced LCOs, we found 4 distinct regions of stimulus parameter space (so called Arnol'd tongues), in which the flies synchronized their wingbeat with the forcing. Furthermore, we found that the stimulation caused a fast modulation of the wingbeat frequency just outside of these parameter regions, demonstrating control of the wingbeat rhythm on a cycle-by-cycle basis. Our results indicate that the direct steering muscles of flies function as a weak mechanical forcing of a limit cycle oscillator embodied in the myogenic wing actuation mechanism.

## 2. Materials & Methods

### 2.1 Flies

We obtained fruit flies (*Drosophila melanogaster* Meigen) from our laboratory stock (descended from a wild-caught population of 200 mated females). A standard breeding procedure was applied (25 females and 10 males, 12:12 hour light/dark cycle, standard nutritive medium). The experiments were performed during the first 8 hours of a subjective day with 5-10 days old female flies.

### 2.2 Experiments

We cold- anaesthetized single fruit flies and glued them by their thorax to the tip a steel tether (Fig. 2A). We then attached the tether to a piezoelectrical actuator (P830.40, Physik Instrumente, Germany). We used a function generator (Agilent 33120A, Hewlett-Packard, USA) to create a sinusoidal voltage signal, amplified it using a custom built amplifier and applied to the piezo. The piezo-induced displacement of the tether was proportional to the voltage with a conversion factor of 0.9 μm/V. The resulting sinusoidal displacement of the tethered fly caused an inertial force $F_i$ on the endknobs of its halteres, given by $F_i(t) = -A_T \omega^2 \sin(\omega t) m_H$, with $A_T$ the amplitude of the tether displacement, $\omega$ the signal's frequency and $m_H$ the mass of the haltere endknob [24].

We oscillated the flies along the anterior-posterior axis, applying either voltage frequency sweeps (0.1-500 Hz, sweep rate 5 Hz/s, holding constant the voltage at 1, 2, 4, 6, 8 or 10 volts) or voltage amplitude sweeps (0.1-8 V, sweep rate 0.8 V/s, holding constant the frequencies close to the baseline wingbeat frequency of the fly currently being tested). Note that as the inertial force is proportional to the square of the stimulus frequency, the forcing amplitude slowly increases during a voltage frequency sweep - despite the voltage amplitude at the piezo staying constant.

When the stimulus frequency $\omega$ is equal to (or very near) the wingbeat frequency, and when the stimulus and the wingbeat are in the appropriate phase relation, the inertial force $F_i$ resembles the Coriolis forces acting on the halteres during pitch manoeuvres. Specifically, to mimic the time course of the Coriolis forces, the phase relation must be such that the tether acceleration is zero at the time points corresponding to the dorsal and ventral wing reversal [28]. The amplitude range of our stimulation (1 - 10 V) then corresponds to pitch rotations of 120 - 1200 °/s (see electronic supplementary material), which lies in the range typical for free flight manoeuvres [42].

We recorded the stimulation and the fly's response together by measuring the velocity of the tether vibration about 1 mm above the fly's body using a laser Doppler vibrometer (MSA-500, Polytec, Germany), sampling at 10240 Hz. We chose appropriate tethers (stainless steel, 1cm length, 200µm diameter) to tune the system suitably for our measurements. In each wing cycle, the wings generate a sharp force peak that is known to occur near the ventral reversal phase of the wingbeat [43]. This leads to an overdamped, high-frequency (>1000 Hz) oscillation of the tether within each wing cycle. In the measured tether velocity trace, this fly-induced signal was overlaid with a low frequency signal corresponding to the piezo actuator (0.1-500 Hz) – see Fig. 2B, top trace.

A

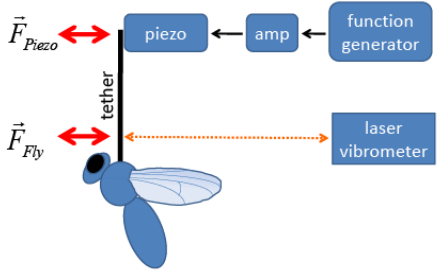

B

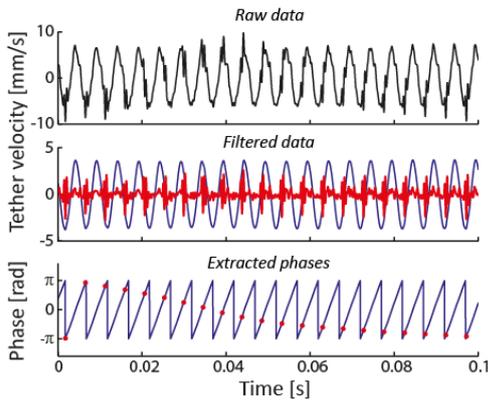

**Figure 2. A) Experimental setup.** Flies were tethered to a steel pin, which was attached to a piezoelectric actuator. The piezo oscillated the tether, and with it the fly, according to the applied voltage signals created by a function generator. In addition, the fly's wingbeat causes the tether to vibrate; we measured the total tether oscillation using a laser Doppler vibrometer. **B) Data processing.** Top panel: Raw data of the tether vibration generated by a flying fly and a sinusoidal piezo oscillation with a frequency close to the baseline WBF of the fly (about 200 Hz). Middle panel: The raw data was high-pass and low-pass filtered to separate the fly's signal (spiky trace, red) from the piezo signal (sinusoidal trace, blue). Bottom panel: Instantaneous phase of the piezo oscillation (blue trace) and the phase in which the main peak in the fly's signal occurred (red dots). **C) Schematic Arnol'd tongues.** Four selected regions of synchronization (filled triangles) for a generic limit cycle oscillator with natural frequency $f_{nat}$, forced by an external stimulus of frequency $f_{stim}$ and amplitude $a_{stim}$. At low stimulus amplitudes, locking of the oscillator to the stimulus occurs when $f_{stim}$ is close to a rational multiple of $f_{nat}$; the frequency range increases with increasing $a_{stim}$. Arrows indicate amplitude and frequency sweeps (see text).

C

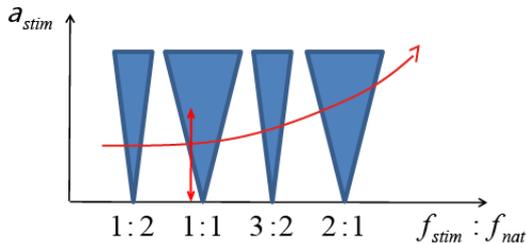

**2.3 Data analysis**

**Phase extraction**

To extract the wing stroke phase from the tether vibration, we applied a 7$^{th}$ order high-pass Butterworth filter in a zero phase lag configuration with a cut-off frequency of 1000 Hz (red trace in Fig. 2B, middle panel). We then up-sampled the data to 50 kHz and applied a peak detection algorithm to determine the times $t_k$ of the prominent peak in this fly-induced signal. The times $t_k$ approximately coincide with the ventral reversal of the wings.

To extract the stimulus phase, we applied a 7$^{th}$ order low-pass Butterworth filter with a cut-off frequency of 500 Hz (blue trace in Fig. 2B, middle panel). We then applied the Hilbert transform to obtain the forcing phase $\Psi(t)$ (blue trace in Fig. 2B, bottom panel). Zero phase (i.e., $\Psi = 0$ *modulo* $2\pi$) was chosen to coincide with the time points at which the tether moves with maximal velocity in the direction towards the posterior direction of the fly (Fig 2A). Note that when $\Psi(t) = 0$ and $\Psi(t) = \pm \pi$, the acceleration of the tether becomes zero.

**Synchrogram construction**

We applied a method called *synchrogram analysis* to reveal synchronization between the stimulus and the response in our data. This method was developed and successfully applied to detect phase-locking in nonstationary and irregular signals [44, 45]. A synchrogram can be described as a 'phase stroboscope': Intervals of synchronization between an external forcing and an oscillator are revealed by plotting the phase of the oscillator at periodic instances of the forcing, or *vice versa*.

We constructed the synchrograms by computing the phase $\Psi(t_k)$ of the forcing at the periodic instances of the wingbeat, (red dots, Fig. 2C). When the fly's wing stroke and the applied force are not synchronized, $\Psi(t_k)$ changes rapidly (Fig. 2B, bottom panel, 0 - 0.7s). When the fly phase-locks to the stimulus and beats its wings at the forcing frequency (*1:1* synchronization), $\Psi(t_k)$ remains constant (single, almost horizontal line in Fig. 2B, bottom panel, 0.7 - 1s). In the case of higher-order synchronization (i.e. *n* wingbeats fit to *m stimulus cycles*, *n:m* locking), the phase points form *m* roughly horizontal lines in the synchrogram.

**Automated synchrogram evaluation**

To determine the stimulus parameter regions in which synchronization took place (i.e. the Arnol'd tongues), we adapted a synchronization detection algorithm for synchrograms from [46] and [47]. Because higher-order lockings tend to be difficult to detect [6], we restricted our analysis to four low-order lockings (*1:2*, *1:1*, *3:2* and *2:1*). The synchronization detection algorithm operates as follows on the synchrogram time series $\Psi(t_k)$. First, the phase is rewrapped according to the locking index *m*, such that $\Psi_m(t_k) = \Psi(t_k) \, modulo \, 2\pi m$ in the vicinity of each potential *n:m* locking region. Next, for each data point in the series, a window spanning 10 wingbeat cycles preceding it and following it is defined, and the mean and the standard deviation of $\Psi_m$ in this window are calculated. If the standard deviation is higher than a threshold of 2.5% of $2\pi m$, the data point is deleted. Following this step, the time series is defined on a single or multiple intervals in which the wingbeat is phase locked with the stimulus. Multiple intervals are present if a transient loss of locking occurs while still within the synchronization region (so called *phase slips* [6], see Sec. 3.1 for an example). To determine the width of the synchronization region, we therefore merged multiple intervals by linear interpolation if 1) their length was at least 15 points (wingbeats) and 2) their distance was less than 10 points. The longest remaining interval was considered to be the candidate region of synchronization.

To exclude false positives, i.e. intervals in which an increase in WBF was not caused by the stimulation but by random fluctuations in the wingbeat frequency, the candidate region was classified as "synchronized" only if the total number of points in the interval was above a threshold value $N_{nm}$. To determine $N_{nm}$, we constructed synchrograms from control data measured without piezo stimulation. This control synchrogram was constructed as described above, but in this case the measured phase $\Psi$ of the stimulus was replaced by the phase of a fictitious stimulus. The synchrogram analysis was then run as described in the previous paragraph. We found that by setting $N_{12} = 45$ for *1:2* locking, $N_{11} = 140$ for *1:1* locking, $N_{32} = 117$ for *3:2* locking and $N_{21} = 218$ for *2:1* locking, our algorithm discarded at least 99.5 % of lockings (of each order) in the control synchrogram. Consequently, we used these threshold values of interval length to declare a candidate region as "synchronized" in our test data sets (with piezo stimulation). The first and the last point of the synchronized interval were taken as beginning and end of the respective Arnol'd tongue.

**Construction of Arnol'd tongues**

To place these border points into the frequency–amplitude parameter space of the forcing, we determined the stimulus amplitudes, i.e., the acceleration amplitude of the tether at the beginning and end of the synchronized interval. The acceleration was calculated from the measured velocity signal according to [48]. To account for possible differences in the baseline frequency (i.e., the wingbeat frequency in absence of piezo stimulation) in between tests and in the 6 flies tested; we divided the stimulus frequency at the border points by the fly's mean wingbeat frequency in the 3 seconds preceding the synchronized interval plus the 3 seconds following the synchronized interval.

For each frequency sweep test during which synchronization of a given order occurred, we therefore placed two border points into the parameter space, with the ratio of stimulus frequency to baseline frequency on the abscissa, and the acceleration delivered by the piezo on the ordinate. The set of all such border points (from the 6 flies tested) defines the Arnol'd tongue.

## 3. Results

### 3.1 Nonlinear response to periodic mechanical stimulus

When we stimulated flies periodically with the piezo, we observed nonlinear responses typical for forced limit cycle oscillators. We will first present three examples to describe these phenomena in detail.

Periodically forced limit cycle oscillators are generically expected to exhibit multiple Arnol'd tongues, corresponding to synchronization at various commensurate ratios of stimulation frequency and oscillator frequency. The frequency range in which locking of a given order occurs typically increases with the forcing amplitude and gives rise to the characteristic tongue shape (Fig. 2C). A transition to synchronization should occur when the stimulus parameters are varied so that the border of the tongue is crossed. This can be achieved e.g. by a stimulus amplitude sweep (vertical arrow in Fig. 2C) or stimulus frequency sweep (horizontal arrow in Fig. 2C).

To test for such a transition at the border of the 1:1 tongue (wingbeat frequency equal to stimulus frequency), we stimulated flies with amplitude sweeps at fixed frequencies close to the fly's baseline frequency. For example, we increased the amplitude from 0 to 3.8 m/s at a fixed *detuning* (difference of baseline frequency and stimulus frequency) of about 5 Hz (Fig.3A). Here, the synchrogram reveals synchronization as a line of constant relative phase $\Psi_m(t_k)$ (black dots, top trace), while the stimulus amplitude is above 2.3 m/s$^2$ (red dotted line, bottom trace). This *phase locking* lasts for about 7 seconds or about 1650 wingbeats. It is interrupted by three intervals, during which the phase changes rapidly (arrows in Fig. 3A). The occasional occurrence of such rapid phase changes, called *phase slips*, is typical for biological oscillators due to their inherent noisiness [6]. The calculated wingbeat frequency (WBF, black line, bottom trace) shows that the fly returns to its baseline frequency during the phase slips.

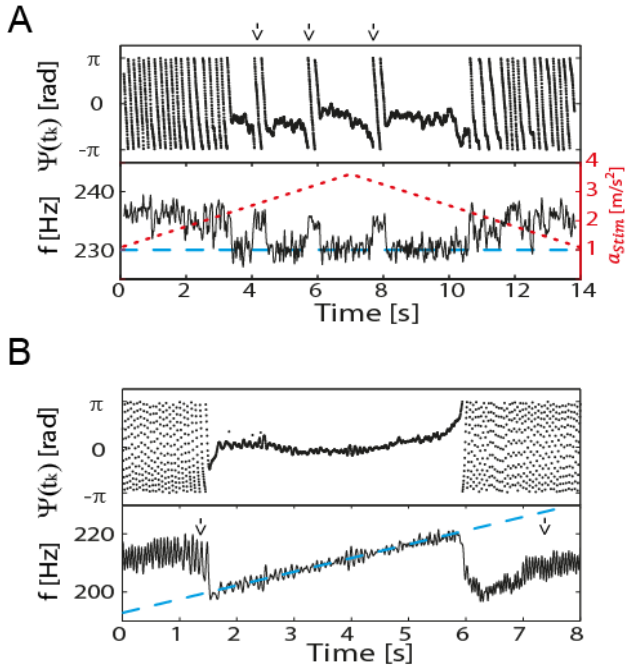

**Figure 3. A) Example of a recorded amplitude sweep.** We stimulated with $f_{stim} = 230$ Hz while linearly first increasing and then decreasing $a_{stim}$, thus crossing the 1:1 Arnol'd tongue border vertically. The fly's baseline WBF was 235 ± 3 Hz. Top: Relative phase $\Psi(t_k)$ between stimulation and wingbeat. Bottom: WBF, $f_{stim}$ (dashed, blue), $a_{stim}$ (dotted, red). Arrows on top indicate phase slips. **B) Example of a recorded frequency sweep.** We stimulated with mean $a_{stim} = 8.1$ m/s$^2$ while linearly increasing $f_{stim}$, thus crossing the 1:1 Arnol'd tongue border horizontally. The fly's baseline frequency was 211 ± 2 Hz. Top: Relative phase $\Psi(t_k)$ between stimulation and wingbeat. Bottom: WBF, $f_{stim}$ (dashed, blue). Arrows indicate quasiperiodic modulation of the WBF.

.

Similarly, we tested for a transition to synchronization when decreasing the detuning. We stimulated flies with frequency sweeps with starting frequencies well below their WBFs. We then increased the stimulus frequency while keeping the applied voltage amplitude constant. In the example shown in figure 3B, the fly's baseline WBF was 209 ± 4 Hz (mean ± std.) and we stimulated with a frequency sweep (0.1-500 Hz, 10 V). The fly locked to the forcing when the stimulation frequency was above 199 Hz and de-locked when it reached 220 Hz. The locking lasted for about 4 seconds or 835 wingbeats. Before and after looking, the WBF oscillated (see arrows). This oscillation reflects a regime in which the forcing frequency is too far away from fly's baseline frequency to fully entrain the fly; however it is close enough to significantly influence the oscillator. Indeed, similar frequency oscillations are generically expected for limit cycle oscillators when the forcing parameters are just outside the Arnol'd tongues (called *phase walk-through* or *quasiperiodicity* [6, 49]). The phenomenon is likewise visible in Fig. 3A just before the onset of locking.

Besides *1:1* entrainment of the oscillator frequency to stimulus frequency, forced limit cycle oscillators typically also show *higher-order synchronization* (also called subharmonic and superharmonic entrainment in the literature). The corresponding Arnol'd tongues touch the abscissa axis (Fig. 2C) at points obtained as rational fractions of stimulus frequency and baseline frequency, $f_{Stim}: f_{Nat} = n: m$, where $n, m$ are relatively prime integers larger than 1. The most robust tongues are generically expected for $n, m$ small (for example *1:2*, *3:2*), while the tongues with large $n, m$ tend to be very narrow and difficult to observe in the presence of noise.

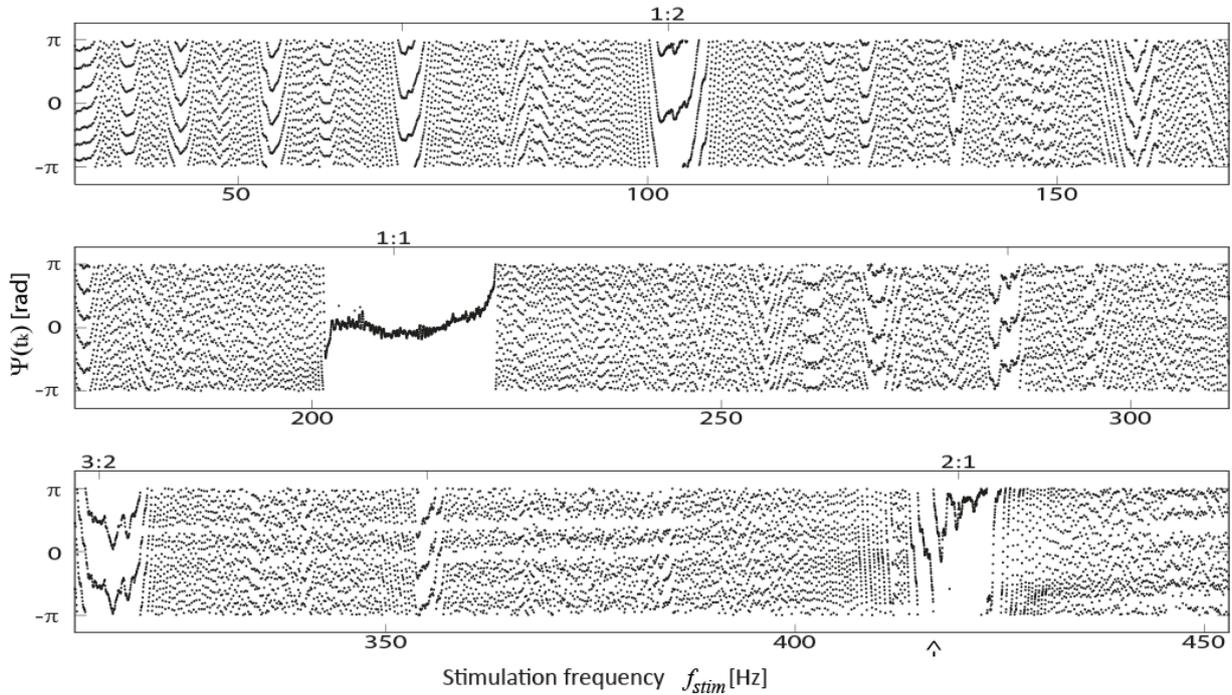

**Figure 4. Synchrogram of a full frequency sweep.** Relative phase $\Psi(t_k)$ between stimulation and wingbeat. We stimulated with a frequency sweep from 0.1 to 500 Hz with 10V applied piezo voltage. The baseline frequency of the fly was 209 ± 4Hz. The continuous recording is shown divided in three rows for clarity. Phase locking leads to the occurrence of horizontal or weakly tilted lines of phase points; for identified lockings of more than 100 wingbeats the corresponding Arnol'd tongue order *n:m* is indicated on top. Arrow indicates phase slip at *2:1* locking.

To test for the existence of higher-order synchronization, we applied frequency sweeps and constructed synchrograms in a range from 40 to 455 Hz. As the baseline wingbeat frequencies were around 200 Hz, the sweeps were expected to cross the borders of multiple Arnol'd tongues. If, for example, a fly beating with 195 Hz is stimulated at 400 Hz, we expect the fly to increase its frequency to 200 Hz, so that the frequency ratio adjusts to the locking order *2:1*. An example synchrogram of a full recording is shown in Fig. 4. Here, the synchrogram reveals not only the extended *1:1* locking region (around 210 Hz), but also several lockings of higher order *n:m*. The fly synchronized in intervals in which the phase points form a single horizontal line (for locking orders with *m=1*, like *1:1* and *2:1*) or *m* horizontal lines (for *m>1*, see also Sec. 2.2). In the presented example, all prominent lockings with a minimum length of 100 wingbeats are marked on top with the corresponding locking order. Note the phase slip at the beginning of the *2:1* locking. This measurement contains 4 prominent lockings, with *3:2* being the highest locking order.

### 3.2 Locking statistics and Arnol'd tongues

To extract the Arnol'd tongue boundaries and to quantify the occurrence of phase-locking, we focused on the frequency sweep experiments. This had the major advantage that by applying a single sweep, we crossed the borders of multiple Arnol'd tongues and, in case of synchronization, obtained both an entry and an exit point of the respective Arnol'd tongue. We tested

6 flies and stimulated each with 6 frequency sweeps of varying voltage amplitude (see Sec. 2.3). We obtained results from 33 measurements (Table 1); during 3 trials, the flies stopped flying and the respective measurements were discarded. *1:1* locking occurred in 16 out of 33 trials (48%). *1:2, 3:2* and *2:1* lockings occurred in 56%, 52% and 33% cases, respectively. We found *1:1* entrainment and higher-order locking in every fly. The percentage of sweeps, in which locking was observed in a fly, was 47 ± 10% (mean ± std.).

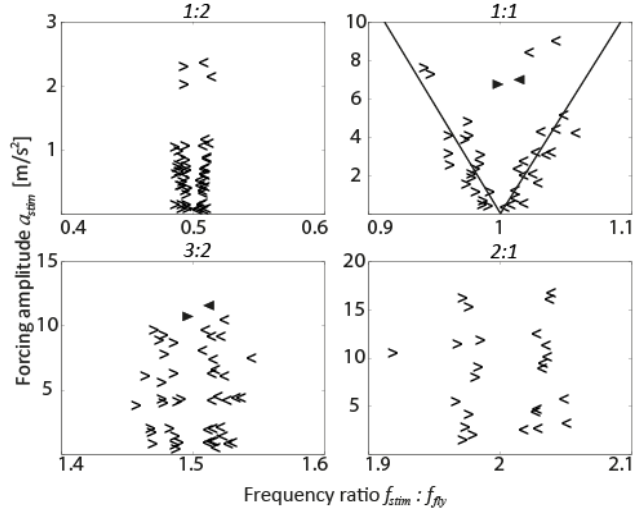

**Figure 5. Extracted Arnol'd tongues.** Start (>) and end points (<) of synchronization as automatically detected from the frequency sweep synchrograms. For the *1:1* lockings we show linear fits of the boundaries of the tongue (robust fits, forced through [1,0]). From the control experiments with flies whose haltere endknobs were ablated, we obtained only two lockings (►◄) with markedly reduced width.

From the identified locking intervals, we reconstructed the Arnol'd tongue boundaries, as described in Sec. 2.3. We found that the *1:1* tongue is roughly triangular, i.e. an increasing forcing strength leads to a roughly proportional increase in locking width, as expected for a weakly forced limit cycle oscillator (Fig. 4, second panel). The width of the tongue shows that the synchronization is maintained for a detuning up to ± 5% of the WBF (which corresponds to about ±10 Hz). The shapes of the higher-order tongues (Fig. 5, panels 1, 3 and 4) are more inconclusive and do not appear to be triangular. Still, all tongues cover extended areas in the frequency-amplitude space.

Note that each measured tongue is nearly symmetrical, which is a generic property of very weakly forced limit cycle oscillators [6]. While it cannot be entirely excluded that a hysteretic effect was nevertheless present due to a frequency scanning in one direction only, it is rather unlikely that this would have exactly compensated an existing intrinsic asymmetric shape. Furthermore, the possible existence of a hysteresis effect does not challenge any of the drawn conclusions in this work.

**Table 1. Locking quantification.** Lockings rates, i.e. percentage of frequency sweeps in which phase locking was observed, obtained from 33 frequency sweeps at piezo voltage amplitudes of 1-10 V.

|   | Locking rate per locking order | | | | Locking rate per fly |
|---|---|---|---|---|---|
|   | *1:2* | *1:1* | *3:2* | *2:1* |   |
| **Fly 1** | 100 | 100 | 83 | 83 | 92 |
| **Fly 2** | 25 | 50 | 50 | 25 | 38 |
| **Fly 3** | 67 | 67 | 67 | 67 | 67 |
| **Fly 4** | 50 | 33 | 33 | 0 | 29 |
| **Fly 5** | 60 | 20 | 60 | 20 | 40 |
| **Fly 6** | 33 | 17 | 17 | 0 | 17 |
| **Mean locking rate per locking order** | 56 | 48 | 52 | 33 | **Mean locking rate per fly ± std.** 47±10 |

### 3.3 Fast and phase-dependent modulation of wingbeat frequency

We showed that sustained mechanical stimulation at a fixed frequency (or slowly increasing frequency in the case of a frequency sweep) acts as a forcing that can entrain the wingbeat rhythm. This result, however, does not tell us the time scale on which the forcing acts. To deduce this time scale, we analysed the intervals of fast WBF modulation prior to *1:1* locking (see Sec. 3.1) in more detail. Such intervals were readily identified for stimulus amplitudes of more than 2 m/s$^2$. In these intervals, the WBF was found to oscillate with a frequency that increased with the difference between baseline WBF and stimulation frequency. We observed WBF oscillations with frequencies up to 40 Hz. In the example shown in Fig. 6A, the time courses of WBF and the phase of stimulation reveal that the WBF varies from cycle to cycle, and follows the phase $\Psi$ of stimulus relative to the wingbeat.

For further analysis, we selected recordings, in which the WBF oscillation lasted for at least 2 seconds (7 measurements from 4 flies). The range of modulation frequencies was 11 to 18 Hz (mean 14Hz, std. 2.5 Hz). We found that the response to stimulus is consistent across the recordings, and that the resulting change in WBF varies approximately sinusoidally with the relative phase $\Psi$ (Fig. 6B). The observed response function (which by definition is periodic in $\Psi$) can be well fitted (r$^2$ = 0.92) with the first two components of its Fourier series (Fig. 6B). A maximal decrease in WBF is observed around relative phase $\pm\pi$, while a phase offset of 0 led to a maximal increase in WBF. Stimulation with phase offsets of $-\pi/2$ and $\pi/2$ did not elicit a response in WBF. Recall that when the stimulus is applied with the frequency close to WBF and with the relative phase $\Psi = 0$ or $\Psi = \pm\pi$, the resulting acceleration has a time course mimicking the Coriolis force acting on the haltere knobs during a pitch manoeuvre in free flight (see Sec. 2.3).

For other values of $\Psi$, the time course of the acceleration does not match the Coriolis force in any rotational manoeuvre during free flight. The response of the wing/thorax oscillator is thus maximal when the stimulus mimics the Coriolis force, to which the haltere mechanosensory pathway is known to be highly sensitive [25].

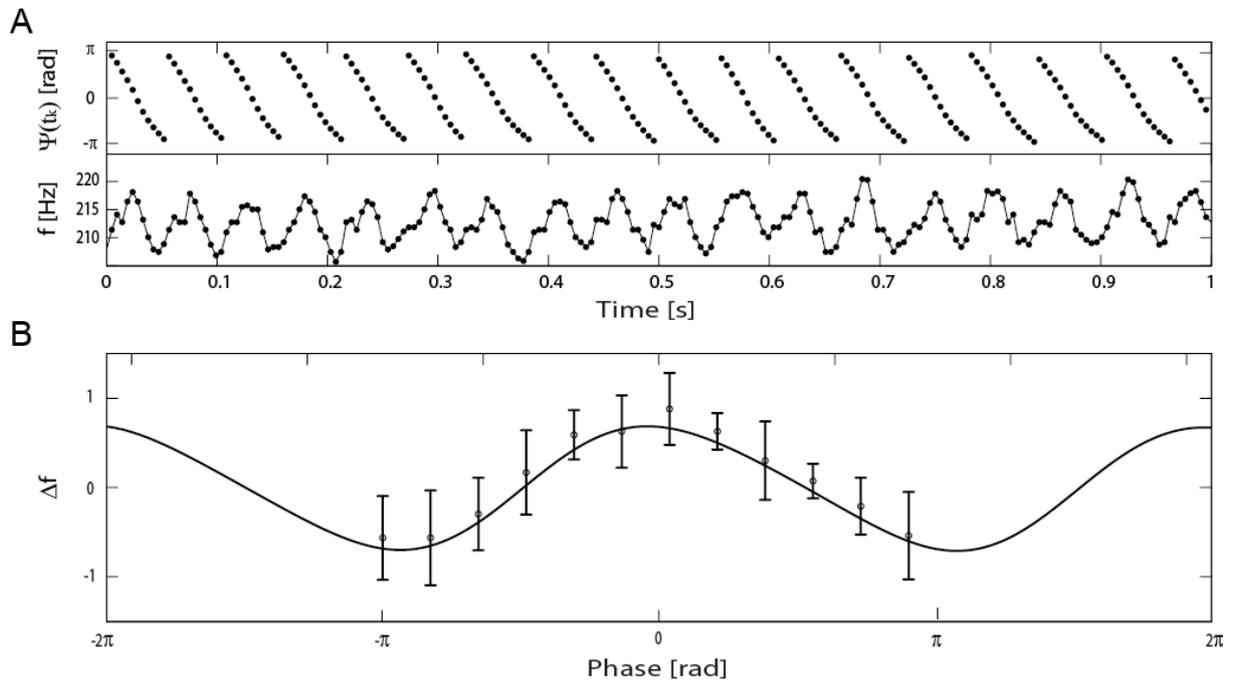

**Figure 6. A) Example of fast modulation of wingbeat frequency during stimulus frequency sweep.** Recording of 1 second of flight prior to the onset of *1:1* locking. Upper trace: Relative phase $\Psi(t_k)$ between stimulation and wingbeat. Lower trace: WBF. **B) Phase-coupled frequency response.** Pooled results from 7 recordings of stimulus frequency sweeps in 4 flies. For each recording, we calculated the difference $\Delta f$ instantaneous WBF and average WBF over the interval [-1.5s:-0.5s] prior to locking, and normalized the time series by its variance. We then binned the data with respect to the stimulus phase. Pooling the responses from the 7 recordings, we calculated the median (open circle) and the interquartile range (error bars). Trace: best fit $y = 0.55\sin(x+1.52) + 0.06\sin(2x-3.1)$ for the two leading Fourier components of the periodic response function.

Based on the observed phase-dependent response, we can make a deduction on how quickly the stimulus affects the WBF. Suppose that a fixed time delay $\tau$ elapses between the mechanical stimulus and the resulting WBF adjustment. In the response curve in Fig. 6B, the time delay $\tau$ would appear as a phase delay $\tau/T$, where $T$ is the period of the modulation of WBF. The response curve was obtained, however, from data segments in which the modulation frequency varied widely (by 60%). Therefore a sizeable time delay $\tau$ would result in a range of phase delays, and after averaging, a clear phase-dependent response (as in Fig. 6B) would likely not be obtained. In addition, the coincidence of the optimal stimulus (resulting in maximal response) with the Coriolis force would be lost. This suggests that the time delay $\tau$ is small compared to the typical period of WBF modulation (about 0.07 sec, or 14 wingbeat cycles). Our analysis therefore indicates that the mechanical stimulus affects the WBF within a few wingbeat cycles.

**3.4 Control experiments**

To test if the observed entrainment was indeed mediated by the halteres, we measured the responses of flies whose haltere endknobs had been ablated (which reduce the haltere sensitivity by about 90% without damaging the sensory fields at the haltere base [25]). We performed 6 frequency sweeps per fly. In 4 out of the 6 flies with ablated halteres, we found no entrainment. In two flies we observed a single, reduced locking region at the highest stimulation strengths (Fig. 5, filled triangles). Our direct observations of the wingbeat of tethered flying flies revealed no gross difference between flies with ablated halteres and untreated flies. Likewise, no qualitative differences were found in the vibrometry traces. Comparison of wingbeat frequencies (extracted from the tether vibration) showed, however, an increased mean baseline WBF ($225 \pm 14$ Hz) in flies with ablated halteres, which was about 12 % elevated, compared to untreated flies ($202 \pm 16$ Hz). A similar observation was made by Dickinson [25], who reported an increase in WBF of about 24% following endknob ablation.

4. **Discussion & Conclusions**

In this study, we explore the possibility that the steering muscles of flies affect the myogenic wingbeat rhythm as a mechanical forcing of a limit cycle oscillator (LCO). Our synchrogram analysis on the phase data of a mechanical stimulus and the wing stroke reveals several characteristic features of a LCO, namely phase-locking, phase-slips, higher-order locking and quasiperiodicity (Sec. 3.1, 3.2). We interpret the measured fast, phase-dependent control of the wingbeat frequency (Sec. 3.3) to result from the activity of the direct steering muscles that function as a mechanical forcing of a LCO embodied in the myogenic wing actuation mechanism. Our data refute the null hypothesis according to which the power-muscle generated rhythm is functionally decoupled from the activity of the direct steering muscles.

We adopted our experimental approach from the seminal studies performed by Nalbach, who vibrated blowflies back-and-forth to emulate the Coriolis forces acting on the halteres of flies during rotations of the body in free flight [24, 28, 50, 51]. While these studies already revealed the basic synchronization phenomenon (*1:1* locking, [50, 51])), they did not systematically explore the possible LCO properties of the flight apparatus. We extended the previous approach by sampling at a high rate the phase relationship between the wing stroke and the mechanical stimulus, using laser vibrometry. Furthermore, we systematically varied the frequency and amplitude of the applied mechanical stimulus to determine the Arnol'd tongues (i.e., regions of synchronization) of the flight control system, and to induce rapid modulation of wingbeat frequency prior to entrainment.

Our results show that the fly responds to mechanical stimulation as a nonlinear oscillator. The essentially nonlinear properties we observed included transitions to / from synchronization, subharmonic / superharmonic entrainment, and phase slips. More specifically, our measurements showed behaviour typical for a nonlinear oscillator with a stable limit cycle [4, 6]. A special type of chaotic oscillator (with narrow strange attractor and a high degree of phase coherence) could have similar synchronization properties (see, e.g., [6], Sec. 10.1.2). Our observations are, however, most straightforwardly understood in the framework of a periodically forced LCO, and we have found no signs of an attractor other than a single limit cycle.

Nalbach interpreted the *1:1* entrainment observed in blowflies as arising from a pitch illusion reflex [50]. Our experimental approach, however, revealed synchronization also at higher frequency ratios (*1:2* and *3:2*), at which the mechanical stimuli can no longer mimic Coriolis forces [28]. Hence, we interpret the observed synchronization as an emergent property of a periodically forced LCO embodied in the myogenic wing actuation mechanism of flies.

Our results are highly consistent with the notion that fast sensory input from the halteres mediates the forcing. First, the maximal response was obtained when the stimulus mimicked the time course of the Coriolis force that would act on the haltere knobs during a pitch manoeuvre (Sec. 3.3). This agrees with previous studies in which blowflies [24, 50] and fruit flies [25] were mechanically stimulated. Second, the responses were strongly reduced after ablating the end-knobs of the halteres (see control experiments described in Sec. 3.4).

In our study, we used observations of rapid phase-dependent modulation of the wingbeat frequency prior to locking to infer that the forcing acts on a time scale comparable to one wingbeat cycle (Sec. 3.3). We attribute the periodic forcing to the mechanical action of direct steering muscles, which have the fast response properties that can explain the measured strict phase relationship between the mechanical stimulus and the frequency response (Fig. 6). The basalar muscle M.b1 could serve this function, as it is known to modulate the wing kinematics according to the phase of its single contractions during successive wing strokes and receives fast monosynaptic input from the halteres [21, 22, 33, 34].

It might be argued that the halteres directly affect the thoracic rhythm mechanically, rather than acting through the mechanosensory system and steering muscles. We evaluate this possibility as follows. We first consider the order of magnitudes of forces exerted on the thorax by the halteres. During *1:1* locking in our experiments, the acceleration reaction force due to the piezo actuation peaks at around $6 \cdot 10^{-8}$ N (see electronic supplementary material). We then compare it with the estimated force exerted by a steering muscle. In the blowfly *Calliphora*, the peak force of the steering muscle M.b1 is of order $10^{-2}$ N and depends substantially on the phase in which it is activated (Fig. 8 in [52]). Given that *Drosophila* is 10 times smaller in length than *Calliphora* and following the scaling laws for insect flight muscles [53], we estimate the forces generated by the homolog M.b1 muscle in *Drosophila* to be $10^3$ = 1000 times smaller than in *Calliphora*. Hence, the forces generated by *Drosophila's* steering muscles (10 μN) exceed the reaction acceleration forces from the halteres (60 nN) by more than 2 orders of magnitude. We therefore conclude that the mechanical forcing of the thoracic rhythm is predominantly effected by the steering muscles.

While the limit cycle properties of central pattern generators (CPGs) are well understood, these are now also revealed in the myogenic wing actuation mechanism of an insect. It is per se not very surprising that a myogenic oscillation can be described as a limit cycle. Our study, however, reveals a coupling of the fly's haltere mechanosensory pathway to the thorax/wing LCO that is functionally significant for flight control. This is evidenced by substantial changes in wingbeat frequency (about ± 10 Hz) measured when the halteres were stimulated by forces in the range of those occurring during typical flight manoeuvres. We concluded that this coupling is effected by the mechanical activity of the steering muscles. In this way, the flight apparatus of the fly is able to avoid the computationally expensive (and comparatively sluggish) neural mechanism of CPGs, and instead replace it with a direct realization of a mechanical limit cycle oscillator. A mechanical system of this type has been recently implemented in a swimming robot by Seo, Chung and Slotine [54]. Myogenic limit cycle – based control provides an elegant conceptual framework in which to understand how flies can achieve extremely fast and precise flight control with minimal neural resources. The fruit fly offers itself as a model system to explore limit cycle – based control mechanisms that minimize the required computational power and operate at high frequencies. Such knowledge could serve the design of biomimetic micro air vehicles (MAVs), which like flies are under severe constraints in terms of mass and power consumption, and therefore depend on highly efficient control strategies [55, 56].

**Acknowledgements**

This work was supported by the Volkswagen foundation (I/80 984 - 986), Swiss National Science Foundation (CR23I2-130111 / 1) and the Czech Republic (AV0Z50110509 and P304/12/G069). We thank Jonathon Howard, Henri Saleh, Vasco Medici and Hannah Haberkern for valuable discussions and support and an anonymous reviewer for insightful critical comments..